\documentclass[10pt,sigconf,nonacm]{acmart}

\AtBeginDocument{%
  }

\setcopyright{none}               
\renewcommand\footnotetextcopyrightpermission[1]{}
\acmBooktitle{}
\acmYear{}

\acmPrice{}
\usepackage{algorithm}
\usepackage{algorithmic}
\usepackage{tikz}
\usetikzlibrary{positioning,shapes.geometric}
\usepackage{subcaption}
\usepackage{enumitem}
\usepackage[most]{tcolorbox}

\usepackage{listings}
\lstdefinelanguage{Verilog}{
  keywords={module, endmodule, input, output, wire, reg, logic, always, begin, end, if, else, case, endcase, assign, parameter},
  sensitive=true,
  morecomment=[l]{//},
  morecomment=[s]{/*}{*/},
  morestring=[b]"
}
\begin{document}
\pagestyle{plain}

\title{LLM4RTL: Tool-Assisted LLM for RTL Generation}

\author{Jing Jin}
\affiliation{%
  \institution{UC Riverside}
  \city{Riverside}
  \state{CA}
  \country{USA}
}
\email{jjin044@ucr.edu}
\author{Robert Chu}
\affiliation{%
  \institution{Futurewei}
  \city{San Jose}
    \state{CA}
  \country{USA}
}
\email{rchu@futurewei.com}
\author{Ning Yan}
\affiliation{%
  \institution{Futurewei}
  \city{San Jose}
    \state{CA}
  \country{USA}
}
\email{ning.yan.uta@gmail.com}

\author{Masood S. Mortazavi}
\affiliation{%
 \institution{Futurewei}
  \city{San Jose}
    \state{CA}
  \country{USA}
 }
\email{masood.mortazavi@futurewei.com}
\begin{abstract}
Large language models (LLMs) have facilitated impressive progress in software engineering, code generation, tooling, and systems.
Concurrently, a significant body of research has developed which explores a growing variety of methods and systems for applying LLMs to hardware and chip design (e.g., systems for RTL code generation based on functional description).
However, when it comes to open Verilog/RTL code-generation, we need high-quality training samples to build specialized and more effective LLM systems through fine-tuning or low-rank adaptation.
Here, we propose a ``judge-renew-check-renew-check'' (JRCRC) pipeline which updates a current public dataset using a hierarchy of state-of-the-art commercial LLM models differing in their costs and capabilities in RTL code generation. 
This approach achieves a cost-effective mechanism for filtering and refining code-generation samples into a higher-quality training dataset. 
Our experiments also identify some common weaknesses of LLMs in rule-based reasoning and logic, and consequently, in RTL code-generation. 
Having identified these weaknesses, we develop an architecture for incorporating pre-processing tools to dynamically assist the LLMs in inferring logical relationships from tabular data formats. 
With our tools-assisted architecture for RTL code generation, we achieve significant overall performance gains in the VerilogEval benchmark and outperform many state-of-the-art methods. Our LLM4RTL system achieves performance comparable to that of GPT-4O using a significantly much smaller LLM. 
\end{abstract}


\maketitle
\section{Introduction}
Large Language Models (LLMs) have shown a notable performance in natural language processing (NLP) benchmarks~\cite{kalyan2021ammussurveytransformerbased, minaee2024llm, chang2024survey}, and LLM-based code generation has noticeably improved~\cite{sun2025surveyneuralcodeintelligence,brown2020languagemodelsfewshotlearners,chen2021evaluatinglargelanguagemodels}.  LLM agents are now widely used in software engineering. However, despite the increasing interest in generating hardware description languages (HDLs), high-quality Verilog codes are scarce, and verifying the correctness and quality of more abundant (often LLM-generated) datasets (e.g., ~\cite{cui2024origen,yang2025haven,nadimi2024pyranet,rtl-coder}) remains quite challenging. 
A substantial body of work has focused on using LLMs for automatic generation of Verilog code from functionality described in text by fine-tuning a small model to achieve better performance than some baselines (e.g., ~\cite{cui2024origen,yang2025haven,liu2024craftrtl,liu2023verilogeval,nadimi2024pyranet,akyash2025rtl++,rtl-coder,pei2024betterv}). 
Tables abound in textual functionality  descriptions (e.g., waveforms, truth tables, Karnaugh maps, etc.) but LLMs have shown significant weaknesses in reasoning over tables~\cite{wolff2025well,cheng2024inductive,cai2024role}. 

The contributions of the present work are as follows:
\begin{itemize}
    \item 
    We construct a cost-effective pipeline to improve existing datasets.
    \item
    We demonstrate how fine-tuning an LLM on these improved datasets leads to improved RTL generation.
    \item 
    We propose tool-based pre-processing to substantially improve an LLM's code-generation performance.
\end{itemize}

\section{Related Work}
VerilogEval is a comprehensive benchmark proposed for evaluating Verilog generation~\cite{liu2023verilogeval} based on tasks originating from HDLBits~\cite{HLDBits}.
 Fine-tuning with only $8$K samples for a few epochs already improves generator performance on VerilogEval~\cite{liu2023verilogeval}.  GPT 3.5 has been used to generate samples for fine-tuning a smaller model to achieve performance comparable to GPT-4~\cite{rtl-coder}. High-quality exemplars have been shown to boost generator performance~\cite{yang2025haven}. Mutimodal combination of textual instructions with control flow and data flow graphs improves performance~\cite{akyash2025rtl++}. OriGen (\cite{cui2024origen}) presents a code-to-code augmentation pipeline for improving code diversity. The code descriptions from open-source RTL code are extracted and then refined using Claude3-Haiku model. Finetuning with the proposed dataset is shown to improve model performance.
To address training data inadequacy, \cite{nadimi2024pyranet} proposes using imperfect data, including samples with syntax errors. the authors argue that this improves diversity and model performance. Their method uses a weighted training scheme based on data quality and curriculum procedure that starts with simpler data. They also release a large public dataset of 692K samples refined using GPT-4o-mini.

Most of the works cited above include the current practice of generating a training dataset using commercial LLMs.
%
%
We modify this strategy. In resource-limited situations, our pipeline enables updating the public dataset in a cost-effective manner employing up-to-date LLMs. At the same time, we preserve any high-quality data generated by the LLMs of previous generations.  

Synthetic samples, which translate the corresponding non-textual representations to truth tables or transition graphs, have been used to enhance the capability of the model to handle logical relations in tabular data~\cite{liu2024craftrtl}. Instead of finetuning the model with such augmented data, we assist existing models with tools that parse tabular data and generate internal logical relations.

Other approaches include generative discriminators to guide the model toward producing the desired output~\cite{pei2024betterv}, an end-to-end framework to optimize the entire workflow~\cite{yao2024rtlrewriter}, and a commercial LLM agent that achieves high performance on the VerilogEval benchmark ~\cite{zhao2024mage}.  In contrast to these, we mainly focus on improving the performance of smaller fine-tuned LLMs. 
\section{Method}
Our empirical approach to fashioning an LLM-based RTL generator consists of some preliminaries followed by three main stages.
\subsection{Preliminaries}
After conducting preliminary experiments using different 7B models (e.g., Qwen/Qwen2.5-Coder-7B, Deepseek-Coder-7B-base-v1.5, StarCoder2-7B, CodeLIama-7B, etc.), we selected DeepSeek-Coder-7B-Instruct-V1.5 because it demonstrated a strong performance.

We used OriGen~\cite{cui2024origen} data as raw material to screen and refine into our fine-tuning dataset.  Origen provides a large, $222$K dataset of aligned (textual description, RTL code) pairs.

We use low rank adaptation (LoRA) technique to fine-tune the model on four A100 GPUs for $3$ epochs using batch size of $8$, learning rate of $5e^{-5}$, warm-up rate of $0.04$, and a weight decay of $0.01$. 
The detailed description part of each data pair is masked during finetuning, with primary emphasis placed on syntactically correct auto-regressive RTL code generation based on summary context.
\begin{figure}[h]
  \centering
    \resizebox{0.47\textwidth}{!}{

\begin{tikzpicture}[
    node distance=1.5cm and 1cm,
    box/.style={rectangle, draw, thick, minimum width=3.2cm, minimum height=2.2cm, text centered, font=\LARGE, align=center, fill=white},
    decision/.style={diamond, draw, thick, minimum width=2.2cm, minimum height=1.2cm, text centered, font=\LARGE, align=center, aspect=1.5, fill=white},
    arrow/.style={thick, ->, >=stealth, blue, line width=3pt},
    label/.style={font=\LARGE, inner sep=5pt}
]

\node[box, fill=gray!15] (origen) {OriGen\\Samples};
\node[box, fill=blue!20, right=of origen] (deepseeK) {DeepSeek-V3\\Labeling};
\node[decision, fill=orange!20, right=of deepseeK] (label) {Label};

\node[box, fill=green!25, above right=1.5cm and 4cm of label] (final) {Final\\Dataset};

\node[box, fill=blue!20, below=2.5cm of label] (renewal) {DeepSeek-V3\\Renewal};
\node[decision, fill=orange!20, right=3cm of renewal] (compile1) {Compilation\\Check};

\node[box, fill=pink!20, below=2cm of compile1] (gpt5) {GPT-5\\Update};
\node[decision, fill=orange!20, right=3cm of gpt5] (compile2) {Compiler\\Check};
\node[box, fill=red!25, below right=1cm and 2.5cm of compile2] (reject) {Rejected};

\draw[arrow] (origen) -- (deepseeK);
\draw[arrow] (deepseeK) -- (label);

\draw[arrow] (label) to[out=45, in=180] node[label, pos=0.3, above left] {True} (final);

\draw[arrow] (label) -- node[label, pos=0.5, right] {False} (renewal);
\draw[arrow] (renewal) -- (compile1);

\draw[arrow] (compile1) to[out=90, in=270] node[label, pos=0.7, right] {Pass} (final);
\draw[arrow] (compile1) -- node[label, pos=0.5, right] {Fail} (gpt5);

\draw[arrow] (gpt5) -- (compile2);

\draw[arrow] (compile2) to[out=90, in=270] node[label, pos=0.3, below right, yshift=20pt] {No Errors} (final);
\draw[arrow] (compile2) -- node[label, pos=0.5, below, xshift=-20pt] {Errors} (reject);

\end{tikzpicture}}
  \caption{Sample Screening and Refinement. 
  }
  \label{fig:pipeline}
\end{figure}
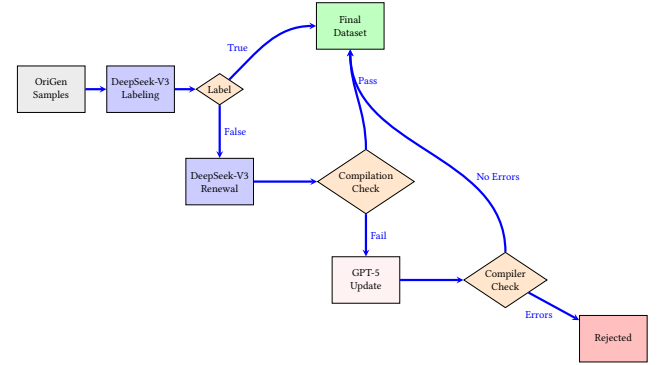
\subsection{Sample Screening and Refinement}
The capabilities of the small models often depend on the commercial LLM models they distill.  The quality of the dataset the commercial models generate directly influences what the smaller models learn. 

Since using a state-of-the-art commercial LLM to refine large quantities of data (e.g., the $222$K samples in the OriGen)  would be a costly proposition,  the earlier well-generated samples (produced by any previous LLM) should be identified and reused. We adopt a nested screen-and-refine pipeline that brings us substantial savings. 

\subsubsection{Data Screening}
We used DeepSeek-V3 (671B), an open-sourced LLM suitable for efficient code generation, as our LLM ``judge'' to screen our data. We substantiated this choice by examining a batch of data samples and discovering a substantial overlap between DeepSeek-V3 and GPT-5 in screening unusable samples. 
Upon this first screening, our judge dismissed $76262$ samples from a total of $222075$ samples. 
\subsubsection{Data Refinement}
Next, we used DeepSeek-V3 to generate solutions to renew these $76262$ samples that were identified as false in the previous stage.  We then compiled the generated solutions with IVerilog to check if any of them had syntax errors. At this stage, $7264$ sample solutions were found to contain syntax errors. We treat these samples as hard cases and now use GPT-5 to generate both solutions and test-benches.  The solutions to around $2000$ hard samples passed the testbench and were considered to be self-consistent.  However, we added all $3361$ renewed samples without syntax errors to the final dataset to increase diversity.  When using GPT-5, we adopted batch-processing to further reduce the API cost.  Figure \ref{fig:pipeline} shows our workflow for updating the dataset.  The refinement cost breakdown (as of September, 2025) was around \$$100$ in DeepSeek-V3 and \$$200$ in GPT-5.   

\subsection{Task Types and Tools}
\subsubsection{Learning Biases}
\label{ss:bias}
 While adding data and improving data quality can help improve overall performance, there exist specific tasks for which simply adding more data may fail to increase \textit{overall} performance of the model (across all tasks) due to biases LLMs exhibit when learning a variety of tasks~\cite{natural_language_reasoning_a_survey}.  
 \begin{figure}[h]
  \centering
  \begin{subfigure}[t]{0.45\columnwidth}
    \centering
 \includegraphics[width=\linewidth]{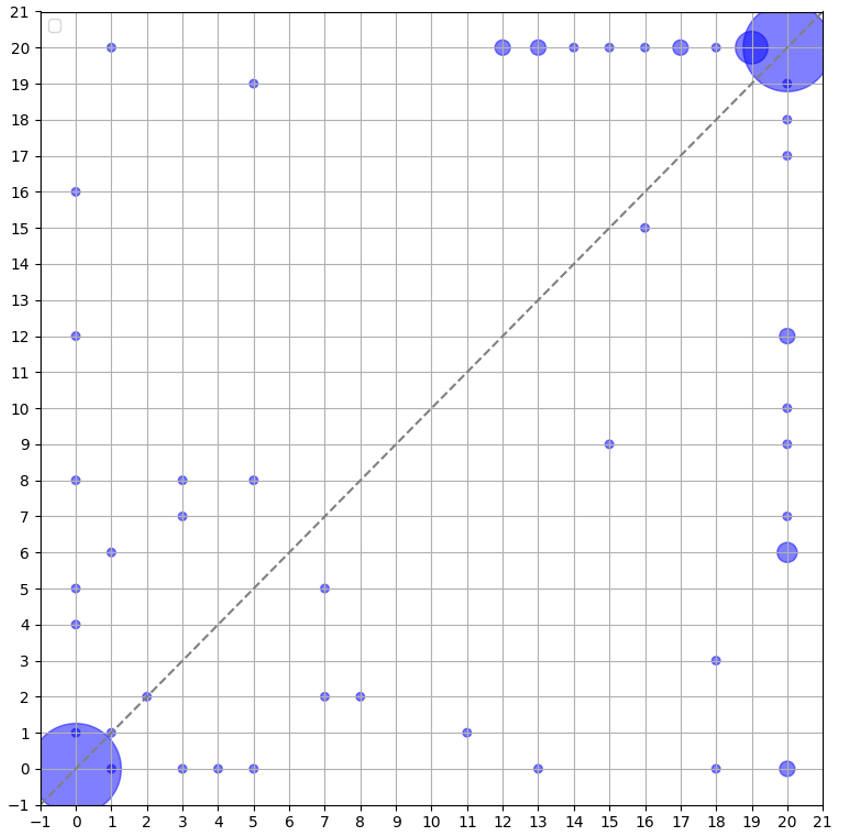}
    \caption{Comparison between the DS-Coder-7b-Instruct-V1.5 ($\mathbf{x}$) and Qwen2.5-Coder-7B-Instruct ($\mathbf{y}$) on updated OriGen dataset. }
    \label{fig:model_compare}
  \end{subfigure}
  \hfill
  \begin{subfigure}[t]{0.45\columnwidth}
    \centering
    \includegraphics[width=\linewidth]{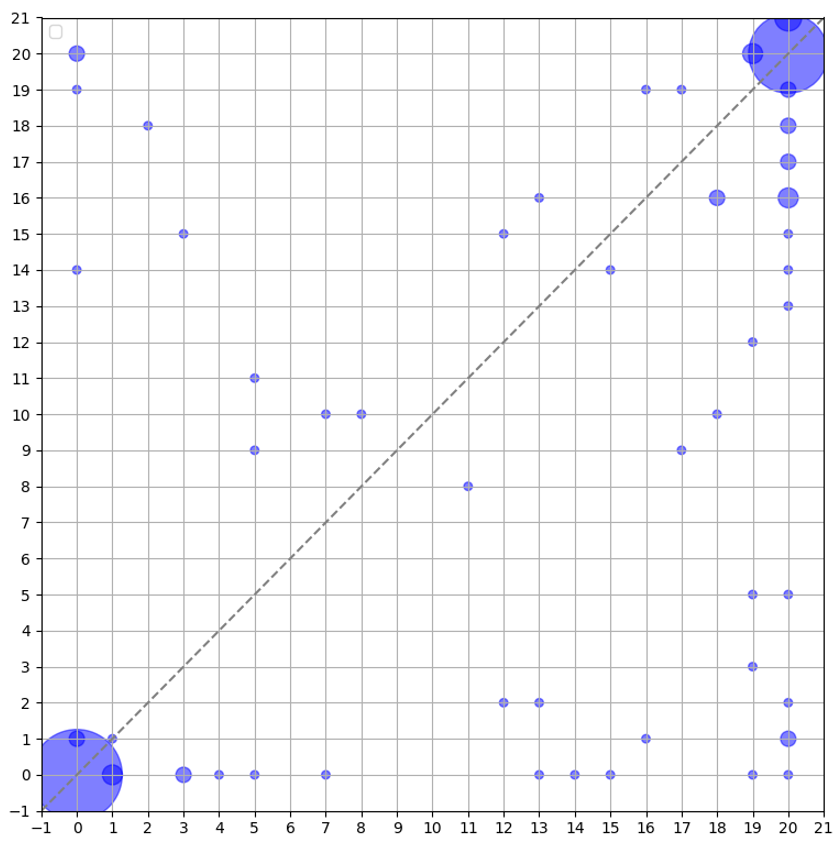}
    \caption{Comparison between DeepSeek-Coder-7b-Instruct-V1.5 finetuned on the dataset OriGen ($\mathbf{x}$) and on the dataset HaVen ($\mathbf{y}$).}
    \label{fig:data_compare}
  \end{subfigure}
  \caption{Comparison on Learning Biases}
  \label{fig:bubble-compare}
\end{figure}
 We illustrate task-mix learning biases in  Figure \ref{fig:bubble-compare}. 
 In Figure \ref{fig:model_compare} we compare two different models (DeepSeek-Coder-7b-Instruct-V1.5 and Qwen2.5-Coder-7B-Instruct), which are both trained on the same dataset (OriGen).  
On the other hand, Figure \ref{fig:data_compare} shows the task-specific performance comparison for the same base model (DeepSeek-Coder-7b-Instruct-V1.5) fine-tuned on two different training datasets, OriGen\cite{cui2024origen} and HaVen\cite{yang2025haven} resulting in two distinct derived models. 
 The point $(x,y)$ in each grid indicates there are $x$ number of successes out of $20$ trials for the model represented on the $\mathbf{x}$ axis and $y$ number of successes out of $20$ trials for the model represented on the $\mathbf{y}$ axis.  
 The size of the point represents the relative number of distinct RTL generation tasks which have achieved the \textit{same} level of $(x,y)$ performance by the two models.
Figure \ref{fig:data_compare} shows biases when two different training dataset are used for the same base model while Figure \ref{fig:model_compare} shows biases when two complete different models are refined on the same dataset. 
In both cases, we have biases.
 \subsubsection{Identifying and Characterizing the Hard Tasks}
In Figure \ref{fig:bubble-compare}, we also observe that there are a number of tasks clustered at the bottom left corner at the point $(0,0)$. This indicates that both compared models perform poorly on a large group of tasks showing 20 failures out of 20 trials for both cases.  
Upon investigating those tasks, we identified a large class of the hard tasks all of which require the models to interpret tables consisting of $0$s and $1$s as well as variables. The RTL solution would require corresponding logical relationships among the variables.  
 \begin{figure}[t] 
\centering 
\begin{tcolorbox}[
  colback=white!5,
  colframe=white!50,
  fonttitle=\bfseries,
  boxrule=0.5pt,
  coltitle=black,
  left=4pt,right=4pt,top=4pt,bottom=4pt
]
\raggedright
\footnotesize\textbf{Instruction:} Read the waveform and implement the same logic.\\
\vspace{6pt}
\renewcommand{\arraystretch}{1.0}
\scriptsize
\centering
\begin{tabular}{rccccc}
  \toprule
  \textbf{t (ns)} & \textbf{a} & \textbf{b} & \textbf{c} & \textbf{d} & \textbf{q} \\
  \midrule
  0  & 0 & 0 & 0 & 0 & 1 \\
  5  & 0 & 0 & 0 & 0 & 1 \\
  10 & 0 & 0 & 0 & 0 & 1 \\
  15 & 0 & 0 & 0 & 0 & 1 \\
  20 & 0 & 0 & 0 & 1 & 0 \\
  25 & 0 & 0 & 1 & 0 & 0 \\
  30 & 0 & 0 & 1 & 1 & 1 \\
  35 & 0 & 1 & 0 & 0 & 0 \\
  40 & 0 & 1 & 0 & 1 & 1 \\
  45 & 0 & 1 & 1 & 0 & 1 \\
  50 & 0 & 1 & 1 & 1 & 0 \\
  55 & 1 & 0 & 0 & 0 & 0 \\
  60 & 1 & 0 & 0 & 1 & 1 \\
  65 & 1 & 0 & 1 & 0 & 1 \\
  70 & 1 & 0 & 1 & 1 & 0 \\
  75 & 1 & 1 & 0 & 0 & 1 \\
  80 & 1 & 1 & 0 & 1 & 0 \\
  85 & 1 & 1 & 1 & 0 & 0 \\
  90 & 1 & 1 & 1 & 1 & 1 \\
  \bottomrule
\end{tabular}
\end{tcolorbox}
\begin{center}
\begin{minipage}{0.95\linewidth}
\begin{lstlisting}[language=Verilog,  
basicstyle=\ttfamily\scriptsize,linewidth=1.0\linewidth,
]
module circuit2(input a, input b, input c, input d, output q);
  assign q = (~a & ~b & ~c & ~d) |
             (~a & ~b &  c &  d) |
             (~a &  b & ~c &  d) |
             (~a &  b &  c & ~d) |
             ( a & ~b & ~c &  d) |
             ( a & ~b &  c & ~d) |
             ( a &  b & ~c & ~d) |
             ( a &  b &  c &  d);
endmodule
\end{lstlisting}
\end{minipage}
\end{center}
\caption{Instruction and  Waveform along with Canonical Solution for a Combinational Circuit.}
\label{lst:circuit2}
\end{figure}

 For example, Figure \ref{lst:circuit2} is a waveform  where the model is required to read from the waveforms and implement the corresponding combinational circuits. 
These kinds of difficult tasks are also identified in~\cite{liu2024craftrtl}, which categorizes them as non-textual representation and adds different categories of chain-of-thought synthetic data in an attempt to improve generative model performance. 
However, according to recent research (e.g.,\cite{cheng2024inductive,cai2024role}), while LLMs are very good at recognizing patterns from a set of examples, they struggle with the tasks requiring rule-based reasoning. When dealing with the waveform problems or other truth-table-related problems, there are well-known procedures that could be applied to derive the SOP (Sum of Products) or POS (Product of Sums) expressions from tabular descriptions~\cite{computer_arch_riscv}.
For example, we have the SOP procedures as shown in Figure \ref{fig:sop}. 
\begin{figure}[t]
\centering
\begin{minipage}{0.95\linewidth}   
\begin{tcolorbox}[
  colback=gray!5,
  colframe=black!50,
  fonttitle=\bfseries,
  boxrule=0.5pt,
  coltitle=black,
  left=6pt,right=6pt,top=6pt,bottom=6pt
]
\footnotesize
\begin{enumerate}
  \item Identify all rows where the output is \textbf{1}.
  \item For each row, write a \textbf{minterm}.
  \item Combine all minterms using OR ($+$).
  \item The result is the \textbf{SOP expression}.
\end{enumerate}
\end{tcolorbox}
\end{minipage}
\caption{Procedure for Deriving Logic Expression in SOP Form}
\label{fig:sop}
\end{figure}

Similarly, there are procedures for parsing the waveforms, for detecting conflicting rows, for eliminating redundant variables, for simplifying final expressions, etc. 
By combining these rules or procedures, we could construct tools or task-specific Python scripts to infer the logical expressions from various waveforms. 
In contrast, it would be hard for the LLMs, particularly the smaller LLMs, to directly learn those rules from the training examples.
Note that pattern matching becomes even more daunting as the number of variables increases or when there exist redundant variables, redundant rows, varying variable names, and conflicting rows. Based on this analysis, we equipped our code generator with a toolbox.        

\subsubsection{Deploying Pre-Processing Tools}
Given our findings regarding training datasets and model variations shown in Figure \ref{fig:bubble-compare} and considering earlier findings in ~\cite{liu2024craftrtl}, we arrived at an approach that can be more easily adapted to different models without the need for fine-tuning those models on additional synthetic data---a practice that can lead to biased learning as shown earlier in Section \ref{ss:bias}. 
%
To prove our toolbox concept, we aim at two types of tasks.

The first task type, for which we include a tool in the toolbox, involves combinatorial logic circuits. For this type of task, the tool only needs to parse the table row by row to identify if it contains logical relationships. The time signal is not important in this kind of problem since the relationships only depend on the information given in the current state.
The second task type we target is the sequential logic circuit. For this type of task, time signals and clock signals are both important because the information from the previous steps would influence the result in the current step. 
Therefore, this second tool needs to first identify the state variables which potentially carry the information from the previous step according to their behaviors related to the clock signals. The tool will then analyze the logical relationships among other variables, the state variables at the current step and the state variables at the next step. This second tool also needs to detect if there exists combinational logic relationships regardless of the time signals. 
These kinds of tools would then pass the logical relationships to the LLM and the LLM would be responsible for translating and assembling these logical relationships into the final RTL code. The general procedure to identify a potential state variable is provided in Algorithm \ref{alg:state}. 

\begin{algorithm}[t]
\caption{Inferring Logic Type of State Outputs}
\label{alg:state}
\footnotesize
\begin{algorithmic}[1]

\FOR{each output signal $y$}
    \STATE Obtain preliminary classification from \texttt{out\_kind[$y$]}.
    
    \IF{$y$ is a positive-edge flip-flop}
        \STATE Try to infer next-state logic from rising edge behavior.
        \IF{inference fails}
            \STATE Reinterpret $y$ as a transparent-high latch.
            \IF{still invalid}
                \STATE Remove $y$ from state variables; classify as combinational.
            \ENDIF
        \ENDIF

    \ELSIF{$y$ is a negative-edge flip-flop}
        \STATE Try to infer next-state logic from falling edge behavior.
        \IF{inference fails}
            \STATE Reinterpret $y$ as a transparent-low latch.
            \IF{still invalid}
                \STATE Remove $y$ from state variables; classify as combinational.
            \ENDIF
        \ENDIF

    \ELSIF{$y$ is a latch (high or low)}
        \STATE Try to infer latch behavior.
        \IF{inference fails}
            \STATE Remove $y$ from state variables; classify as combinational.
        \ENDIF

    \ELSE
        \STATE Classify $y$ directly as combinational logic.
    \ENDIF
\ENDFOR

\end{algorithmic}
\end{algorithm}

We use a similar approach to construct tools for parsing the Karnaugh map. Note that Karnaugh map is another non-texual format to represent a truth table.
%
\subsection{Tool Selection and Iterative Inference Process}
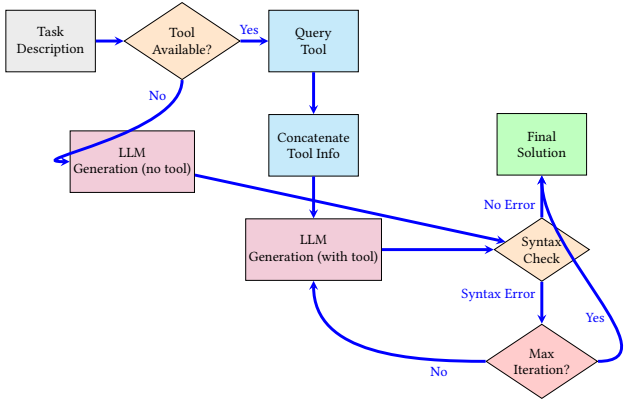
\begin{figure}[htbp]
\centering
\resizebox{0.5\textwidth}{!}{\begin{tikzpicture}[
    node distance=1.5cm and 1cm,
    box/.style={rectangle, draw, thick, minimum width=3.2cm, minimum height=2.2cm, text centered, font=\LARGE, align=center},
    decision/.style={diamond, draw, thick, minimum width=2.2cm, minimum height=1.2cm, text centered, font=\LARGE, align=center, aspect=1.5},
    arrow/.style={thick, ->, >=stealth, blue, line width=3pt},
    label/.style={font=\LARGE, inner sep=5pt}
]

\node[box, fill=gray!15] (start) {Task\\Description};
\node[decision, fill=orange!20, right=of start] (toolcheck) {Tool\\Available?};
\node[box, fill=cyan!20, right=of toolcheck] (tool) {Query\\Tool};

\node[box, fill=cyan!20, below=of tool] (concat) {Concatenate\\Tool Info};

\node[box, fill=purple!20, below=of concat] (llm_enhanced) {LLM\\Generation (with tool)};
\node[box, fill=purple!20, below left=2.5cm and -1.5cm of toolcheck] (llm_direct) {LLM\\Generation (no tool)};

\node[decision, fill=orange!20, right=4cm of llm_enhanced] (compiler) {Syntax\\Check};
\node[box, fill=green!25, above=of compiler] (final) {Final\\Solution};
\node[decision, fill=red!20, below=of compiler] (maxiter) {Max\\Iteration?};

\draw[arrow] (start) -- (toolcheck);
\draw[arrow] (toolcheck) to[out=0, in=180] node[label, pos=0.3, above] {Yes} (tool);
\draw[arrow] (toolcheck) to[out=270, in=180] node[label, pos=0.1, left, xshift=-5] {No} (llm_direct);

\draw[arrow] (tool) -- (concat);
\draw[arrow] (concat) -- (llm_enhanced);

\draw[arrow] (llm_enhanced) -- (compiler);
\draw[arrow] (llm_direct) -- (compiler);

\draw[arrow] (compiler) to[out=90, in=270] node[label, pos=0.3, left] {No Error} (final);
\draw[arrow] (compiler) to[out=270, in=90] node[label, pos=0.3, left] {Syntax Error} (maxiter);
\draw[arrow] (maxiter) to[out=180, in=270] node[label, pos=0.2, below] {No} (llm_enhanced);
\draw[arrow] (maxiter) to[out=0, in=270] node[label, pos=0.4, left] {Yes} (final);

\end{tikzpicture}}
\caption{Tool-Enhanced LLM Code Generation Workflow}
\label{fig:tool-workflow}
\end{figure}

Finally, in Figure \ref{fig:tool-workflow}, we show how  small code-generating LLM models are enabled with the toolbox that we have provided. First, from the given task description, the coordinating program needs to determine which tool to use. Currently, we use a simple way relying on keywords in the problem description to choose the right tool, but this tool selection task can potentially be delegated to another LLM agent or some other more mature mechanism. Second, after choosing the correct tool, the tool will parse the description and possibly also the module header to determine any logical relationships.
The logical information the tool has derived will be concatenated to the task description and passed together to the code-generating LLM. 
Next, the LLM will generate the final solution.
To improve robustness, we add a final iterative step: if the generated solution has syntax errors, we regenerate it until it no longer contains syntax errors or we reach the maximum  iteration limit.
\section{Evaluation}
We use VerilogEval-human (of VerilogEval \cite{liu2023verilogeval}) as the benchmark in all our evaluation experiments. This benchmark contains $156$ different tasks. In each sample, this benchmark provides the task description as well as the module header. It expects the model to output the correct Verilog solution.
We use PASS@1 and PASS@5 as metrics to evaluate  the performance of the model.
\begin{equation}
\label{eq:passk}
\mathrm{PASS@}k
= \mathbb{E}\!\left[\,1 - \frac{\binom{n-c}{k}}{\binom{n}{k}} \right]
\end{equation}
PASS@k (Eq.~\eqref{eq:passk}) expresses the probability that at least one of the top-k generated samples is correct. We generate $n = 20$ samples for each task, with the generative temperature set to $0.5$.

We evaluate several open-source $7$B LLM coders directly on the benchmark. As shown in Table \ref{tab:base-results}, they already show some ability to generate correct Verilog solutions.

\begin{table}[h]
\centering
\caption{Pass Rate with Different $7$B Base Models.}
\label{tab:base-results}
\footnotesize
\begin{tabular}{p{4cm}ccc}
\toprule
Base Model & PASS@1 & PASS@5 \\
\midrule
DS-Coder-7B-Instruct & 0.3372 & 0.4194 \\
DS-Coder-7B-Instruct-V1.5 & 0.3458 & 0.4479 \\
Qwen2.5-Coder-7B-Instruct & 0.3631 & 0.4839 \\
CodeLlama-7B & 0.2058 & 0.3436\\
StarCoder2-7B & 0.0978 & 0.2485\\
\bottomrule
\end{tabular}
\end{table}

\subsection{Step-by-Step Updating of OriGen Dataset}

The experimental results with the OriGen dataset, the semi-refined OriGen dataset (DeepSeek-V3-updated) and the fully-refined dataset are given in Table \ref{tab:data-results}.
These results indicate that our screen-refine---``judge-
renew-check-renew-check'' (JRCRC)---steps improve overall performance and that the pipeline
strikes a good balance between cost and effectiveness. Hence, we have demonstrated how to update
a dataset under limited resources, using newer teacher LLMs while still preserving the many 
high-quality samples produced by earlier LLMs.

%
In our experiments, we only keep training samples shorter than 1,024 tokens. Interestingly, our fully refined dataset is the smallest, yet it delivers the best performance. Only about 3K samples
are updated in the last stage, and
another 3K are removed due to syntax
errors. Our finding here shows that the model is
highly sensitive to data quality---even
a small amount of erroneous code can damage performance.

\begin{table}[h]
\centering
\caption{Pass Rate with Dataset in Three Stages.}
\label{tab:data-results}
\footnotesize
\begin{tabular}{p{3.6cm}ccc}
\toprule
Dataset & Sample size & PASS@1 & PASS@5 \\
\midrule
OriGen  & 174,971 & 0.4952 & 0.5712 \\
OriGen updated with DS-V3 & 168,787 & 0.5138 & 0.5763 \\
OriGen updated wth DS-V3 and GPT-5 & 165,916 & 0.5279 & 0.5971 \\
\bottomrule
\end{tabular}
\end{table}
\subsection{Improving Results Using Tools and Iterations}
\begin{table*}[h]
\centering
\caption{Performance Comparison on  VerilogEval-human Benchmark.}
\label{tab:benchmark}
\footnotesize
\begin{tabular}{p{3.2cm}p{3.6cm}ccc}
\toprule
Training Dataset  & Base Model & Number of Training Samples & PASS@1 & PASS@5 \\
\midrule
- & GPT-4O & - & 0.6022 & 0.6605\\
- & Claude3-Haiku & - & 0.475 & 0.577\\
PyraNet\cite{nadimi2024pyranet} & DS-Coder-7B-Instruct-V1.5 & 692K & 0.583 & 0.628\\
RTL++\cite{akyash2025rtl++} & Codellama-7B & 200K & 0.599 & 0.688\\
CraftRTL\cite{liu2024craftrtl} & DS-Coder-7B & 80K & 0.631 & 0.678\\
RTL-Coder\cite{rtl-coder} & DS-Coder-6.7B & 27k & 0.416 & 0501 \\
OriGen\cite{cui2024origen} &  DS-Coder-7B-Instruct & 180K & 0.544 & 0.601\\
BetterV\cite{pei2024betterv} & CodeQwen1.5-7B-Chat & - & 0.461 & 0.531 \\
Updated Dataset  & DS-Coder-7B-Instruct-V1.5 & -
&0.5279 & 0.5971 \\
Updated Dataset + Tools & DS-Coder-7B-Instruct-V1.5 & - 
&0.6003 & 0.6632 \\
Updated Dataset + Tools + Iterative inference & DS-Coder-7B-Instruct-V1.5 & -
&0.6080 & 0.6673 \\
\bottomrule
\end{tabular}
\end{table*}

\begin{table*}
\centering
\caption{Success Counts for Different Circuit Tasks in VerilogEval-human Benchmark.}
\label{tab:circuit-results}
\footnotesize
\begin{tabular}{ccccccc}
\toprule
Task ID &  Category & Logical-Expression-Related & Our Model & Our Model with Tools & GPT-5 & DeepSeek-V3-671B  \\
\midrule
Circuit1 & Combinational & Y & 9 & 20 & 20 & 20 \\
Circuit2 & Combinational & Y & 0 & 20 & 20 & 20 \\
Circuit3 & Combinational & Y & 0 & 20 & 19 & 0 \\
Circuit4 & Combinational & Y & 0 & 20 & 20 & 0 \\
Circuit5 & Combinational & N & 0 & 0 & 20 & 20 \\
Circuit6 & Combinational & N & 15 & 16 & 20 & 20 \\
Circuit7 & Sequential & Y & 0 & 20 & 1 & 19 \\
Circuit8 & Sequential & Y & 0 & 9 & 20 & 0 \\
Circuit9 & Sequential & N & 0 & 0 & 9 & 19 \\
Circuit10 & Sequential & Y & 0 & 20 & 9 & 0 \\
\bottomrule
\end{tabular}
\end{table*}

%
Besides improving the training data,
we also evaluate the performance gains
from using tools and iterative 
inference (Table \ref{tab:benchmark}). We compare our
model with earlier reports and with
commercial models. Using tools 
boosts PASS@1 from $0.5279$ to $0.6003$ without any retraining. This means the technique can be applied to any model to enhance its performance on specific tasks. 
The iterative inference shows little improvement, likely because most of our training data is already syntax-error-free, making the model's outputs robust to syntax errors. 

With the toolbox, our model reaches
performance comparable to state-of-the-art systems. It matches GPT-4O and outperforms some models trained on much larger datasets. The results also show that high-quality training data is crucial for strong performance.

Moreover, the large overall performance increase is based on tools related to logical expressions. For instance, we list the model performance on a series of  circuit problems in Table \ref{tab:circuit-results}. All these hard problems require models to interpret the waveforms and to translate them to the corresponding Verilog solutions.

\section{Conclusion and Discussion}
In this work, we propose a method to update datasets by using a state-of-the-art commercial language models in a cost-effective manner and to develop tools to assist low-cost models to handle problems with logical and truth-table descriptions. 

We keep all renewed data without syntax errors,
but since LLMs can generate self-consistent solutions and testbenches, adding samples that 
also pass these tests could further improve data quality. Exploring better methods to produce higher-quality data is left for future work.

On the other hand, we develop and use a toolbox to address LLMs' known weakness in rule-based deductions given tabular data. Our current toolbox handles simple circuit modules. In the future, it could be expanded to apply to more complex code structures and task types. Finally, while we used key words in the problem description to decide which tool to deploy, other approaches for tool selection would also be an interesting problem for further investigation. One can potentially use reinforcement learning to optimize a tool selection policy. 

\section{Content Generated by AI}
 We used ChatGPT and Gemini during the preparation of this manuscript. The system was used to improve grammar and clarity, provide suggestions for formatting algorithm boxes and Verilog code examples, assisting in debugging and coding. All generated material was manually reviewed, corrected, and validated by the authors. The model did not generate or modify the research ideas, scientific contributions, analysis, experimental results, or conclusions. The authors assume full responsibility for the accuracy and originality of all content.

\begin{acks}
This work was conducted during the first author's internship at Futurewei Technologies. We thank the Machine Learning Group and the CPU Team of the IC Lab at Futurewei for valuable discussions and support.
\end{acks}

\clearpage
\bibliographystyle{ACM-Reference-Format}
\bibliography{main_ref}

%

\end{document}